\documentclass[letterpaper,conference]{IEEEtran}
\usepackage[utf8]{inputenc}
\usepackage[english]{babel}
\usepackage{cancel}
\usepackage[numbers,sort]{natbib}
\usepackage[dvipsnames]{xcolor}
\usepackage{amsmath}
\usepackage{url}
\usepackage{multirow}
\usepackage{array,multirow}
\usepackage{algorithm}
\usepackage{algorithmic}
\usepackage{graphicx}
\usepackage{verbatim}
\graphicspath{ {./figure/} }
\addto\captionsenglish{}
\usepackage{babel}
\addto\captionsenglish{}
\usepackage{xcolor}
\usepackage{tabularx}
\usepackage{amssymb}
\usepackage{amsthm}
\usepackage{fancyhdr}

\usepackage{array, boldline, makecell, booktabs}

\begin{document}

\title{Maintenance of Structural Hole Spanners in Dynamic Networks}

\makeatletter
\newcommand{\linebreakand}{%
  \end{@IEEEauthorhalign}
  \hfill\mbox{}\par
  \mbox{}\hfill\begin{@IEEEauthorhalign}
}
\makeatother

\author{
  \IEEEauthorblockN{Diksha Goel}
  \IEEEauthorblockA{\textit{School of Computer Science}\\
    \textit{University of Adelaide, Australia}\\
   diksha.goel@adelaide.edu.au}
  \and
  \IEEEauthorblockN{Hong Shen}
  \IEEEauthorblockA{\textit{School of Computer Science and Engineering}\\
    \textit{Sun Yat-sen University, Guangzhou, China}\\
    shenh3@mail.sysu.edu.cn}
  \linebreakand 
  \IEEEauthorblockN{Hui Tian}
  \IEEEauthorblockA{\textit{School of Information and Communication Technology}\\
  \textit{Griffith University, Australia}\\
  hui.tian@griffith.edu.au}
  \and
  \IEEEauthorblockN{Mingyu Guo}
  \IEEEauthorblockA{\textit{School of Computer Science}\\
  \textit{University of Adelaide, Australia}\\
  mingyu.guo@adelaide.edu.au}
    
}

\IEEEtitleabstractindextext{%
\begin{abstract}
Structural Hole (SH) spanners are the set of users who bridge different groups of users and are vital in numerous applications. Despite their importance, existing work for identifying SH spanners focuses only on static networks. However, real-world networks are highly dynamic where the underlying structure of the network evolves continuously. Consequently, we study SH spanner problem for dynamic networks. We propose an efficient solution for updating SH spanners in dynamic networks. Our solution reuses the information obtained during the initial runs of the static algorithm and avoids the recomputations for the nodes unaffected by the updates. Experimental results show that the proposed solution achieves a minimum speedup of 3.24 over recomputation. To the best of our knowledge, this is the first attempt to address the problem of maintaining SH spanners in dynamic networks.\\\end{abstract}

}

\maketitle
\IEEEdisplaynontitleabstractindextext
\IEEEpeerreviewmaketitle

\section{Introduction}
\IEEEPARstart{O}ver the past few years, various large-scale networks have emerged, such as collaboration networks, social networks, etc. The topological structure of these networks exhibits a community structure where the nodes are tightly connected within the community. The presence of community structure in the network leads to the formation of structural holes, which are the gaps formed due to lack of connectivity between the communities \cite{lou2013mining}. It has been shown that the information circulated within the community tends to be similar. Therefore, the presence of SH between the communities leads to the redundant flow of information within the community. Hence, a community needs to have connectivity with different communities to access novel information. SH spanners are those individuals who fill structural holes by acting as a bridge between different communities that are otherwise disconnected \cite{lou2013mining}. Figure \ref{fig:SHS} illustrates the SH spanners in the network. 
SH spanners have numerous real-world applications such as information diffusion, preventing the spread of rumours, community detection, etc. For example, identifying SH spanners in the network and filtering the information passing through them can prevent the spread of rumors in other communities. Several studies \cite{lou2013mining, rezvani2015identifying, gong2019identifying, goyal2007structural, he2016joint, xu2017efficient, xu2019identifying, tang2012inferring} have been conducted for discovering SH spanners in static networks. However, real-world networks are highly dynamic and change rapidly. As a result, identified SH spanners change, and therefore, it is crucial to track updated spanners in the evolving networks. We study SH spanner problem for the dynamic networks. We define SH spanner as a node whose removal minimizes the pairwise connectivity of the residual network. This definition aims to capture the nodes that are located between otherwise disconnected groups of nodes. While the traditional SH spanner problem focuses on discovering the spanner nodes that minimize the pairwise connectivity of the network, the SH spanner tracking problem aims to update the previously identified spanner nodes as the network evolves. This paper proposes an efficient solution that maintains top-$k$ SH spanners for decremental edge updates in the network. Our solution performs greedy exchange, each time replacing an old spanner node with a high score node from the network.

 \begin{figure}[!t]
  \centering
    \includegraphics[width=0.75\columnwidth]{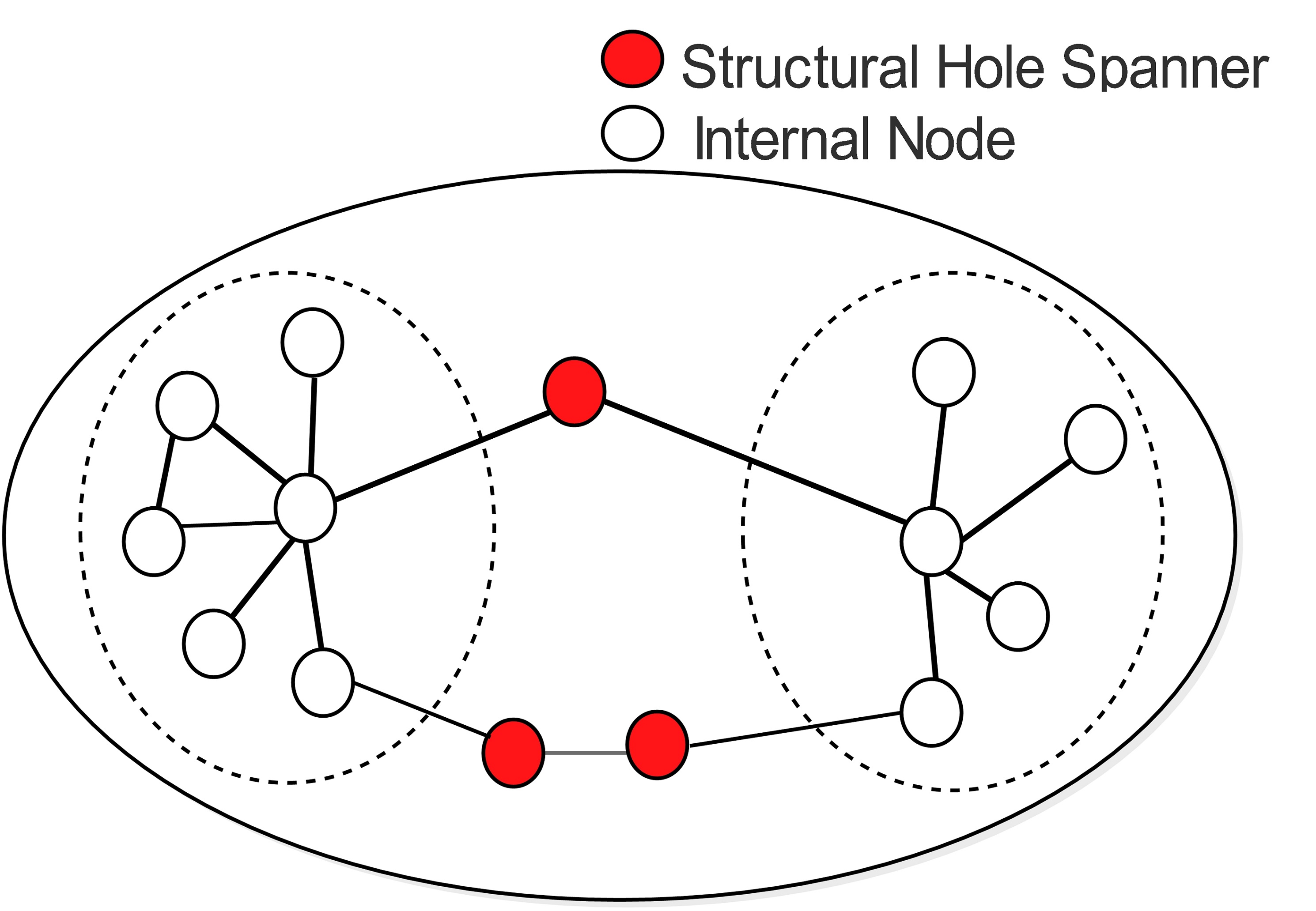}
    \caption{Illustration of structural hole spanners.}
    \label{fig:SHS}
  \end{figure}
  
Our contributions can be summarized as follows. First, we study SH spanner problem for dynamic networks and formulate Structural hole Spanner Tracking (SST) problem. We then propose an efficient solution for SST problem that maintains spanner nodes for single decremental edge updates in the network by discovering a set of affected nodes. We also design a method to compute the pairwise connectivity score of the nodes efficiently. We evaluate the performance of the proposed solution by conducting extensive experiments, and the results demonstrate that our solution achieves a minimum speedup of 3.24 over recomputation.

\section{RELATED WORK}
SH theory was first studied by Burt \cite{burt2009structural} to identify the critical personnel in the company to integrate various operations. Goyal et al. \cite{goyal2007structural} designed a model to illustrate how a node act as a SH spanner in the network. They formulated the SH spanner problem as a set of nodes that pass through numerous shortest paths between distinct pair of nodes. Tang et al. \cite{tang2012inferring} designed a 2-step mechanism to identify SH spanners. For every node, the model only considers the shortest paths having length two, on which the node resides, and the rest of the paths are ignored. Rezvani et al. \cite{rezvani2015identifying} designed several heuristics and argued that eliminating those nodes that bridge multiple communities results in an increase in the sum of all-pair shortest distance in the network. Based on \cite{rezvani2015identifying}, Xu et al. \cite{xu2017efficient} designed fast and scalable algorithms for identifying spanner nodes. Gong et al. \cite{gong2019identifying} designed a machine learning model to discover SH spanners in the online social network. Burt \cite{ burt2011structural} studied the correlation between the strength of the links with which a node is connected to its bridged communities and the bridging advantage of that node. Based on \cite{burt2011structural}, Xu et al. \cite{xu2019identifying} designed maxBlock algorithm to discover SH spanners that connect many communities and have substantial relations with these communities. Lou et al. \cite{lou2013mining} designed a model to discover SH spanners and argued that eliminating the SH spanners from the network leads to a decrease in the minimal cut of the communities. The model requires prior community information however, community identification is an expensive process. He et al. \cite{he2016joint} designed a model that jointly discovers communities and SH spanners.

\section{PRELIMINARIES AND PROBLEM STATEMENT}
\subsection{Network Model}
A network can be modeled as an undirected graph\footnote{We consider only graphs without self-loops or multiple edges.} $G = (V, E)$, where $V$ and $E$ are the set of nodes and edges in the network. Let $n=|V|$ and $m=|E|$ denotes the number of nodes and edges in the graph, respectively. A \textit{path} $p_{ij}$ from node $i$ to $j$ in an undirected graph $G$ is a sequence of nodes ${\{v_{i},v_{i+1},...,v_{j}\}}$  such that each pair  $(v_{i},v_{i+1})$ is an edge in $E$. A pair of nodes $i$, $j \in V$ is \textit{connected} if there is a path between $i$ and $j$. A \textit{connected component} or \textit{component} $C$ in an undirected graph $G$ is a maximal set of nodes in which a path connects each pair of node. The \textit{pairwise connectivity} $u(i,j)$ for any node pair $(i,j)\in V \times V $ is quantified as $1$ if node $i$ and $j$ are connected, and $0$ otherwise.

\noindent \textit{Total pairwise connectivity} $P(G)$, i.e., pairwise connectivity across all node pairs in the graph, is given by:

{\begin{equation}
\text{\ensuremath{{\displaystyle P(G)=\sum_{i,j\in V\times V,i\neq j}{\textstyle u(i,j)}}}}
\end{equation}}

\noindent \textit{Pairwise connectivity score} $c(i)$ of node $i$ is the contribution of node $i$ to the total pairwise connectivity score of the graph. Pairwise connectivity score $c(i)$ of node $i$ is calculated as follows:
{\begin{equation}
c(i)=P(G)-P(G\backslash\{i\})
\end{equation}}

\subsection{Structural Hole Spanner Problem}
\noindent \textbf{Structural Hole Spanner Problem.} Given a graph $G = (V,E)$, and a positive integer $k$, \textit{SH spanner problem} is to identify a set of spanners Top-$k$ in $G(V, E)$, where Top-$k$ $\subset V$ and $|$Top-$k|= k$, such that the removal of nodes in Top-$k$ from $G$ minimizes the total pairwise connectivity in the residual subgraph $ G(V\backslash $Top-$k)$.

\begin{equation}
\text{Top-}k = {min}\,\,\{{P(G\backslash \text{Top-}k)}\}
\end{equation}
where Top-$k\,\subset\,V$ and $|$Top-$k|=k$.

\begin{algorithm}[h]
 \caption{Structural hole spanner identification.}
 \label{alg 1}
 \begin{algorithmic}[1]
 \renewcommand{\algorithmicrequire}{\textbf{Input:}}
 \renewcommand{\algorithmicensure}{\textbf{Output:}}
 \REQUIRE Graph $G(V, E)$, $k$
 \ENSURE  top-$k$ spanner set Top-$k$
  \STATE Initialize Top-$k=\phi$
  \WHILE{$|$Top-$k|<k$}
  \STATE $v'=argmax_{v\,\text{\ensuremath{\in}}\,V}\,c(v)$
  \STATE $G=G\backslash\{v'\}$
  \STATE Top-$k=$Top-$k\bigcup\{v'\}$
  \ENDWHILE
 \RETURN Top-$k$
 \end{algorithmic}
 \end{algorithm}

Algorithm \ref{alg 1} presents a heuristic approach for discovering the SH spanners in the static network. The algorithm works by repeatedly selecting a node $v'$ with a maximum pairwise connectivity score,  which minimizes the total pairwise connectivity of the residual network when removed from the network.

\subsection{Structural Hole Spanner Tracking Problem}
\noindent \textbf{Structural Hole Spanner Tracking (SST) Problem.} Given a graph $G = (V, E)$, spanner set Top-$k$, and edge update $\Delta E$, \textit{SST problem} is to identify a spanner set Top$'$-$k$ with cardinality $k$ in $G'(V,E + \Delta E)$ by updating Top-$k$ such that the removal of nodes in Top$'$-$k$ from $G'$ minimizes the total pairwise connectivity in the residual subgraph $G'(V\backslash$Top$'$-$k$).

A naive solution for SST problem is to apply algorithm \ref{alg 1} after every update, which will return the new SH spanner set. However, this solution is computationally expensive. This paper focuses on updating top-$k$ spanners without explicitly running algorithm \ref{alg 1} after every update.

\section{PROPOSED SOLUTION}
We propose an efficient solution that maintains top-$k$ spanners in the dynamic network. Instead of constructing the spanner set from the ground, we start from old Top-$k$ and repeatedly update it.

\subsection{Finding Affected Nodes}
Whenever there is an edge update in the network, by identifying the set of affected nodes, we need to recompute the scores of these nodes only. When an edge $(x,y)$ is deleted from the network, affected nodes $A$ are the nodes reachable from node $x$, in case edge $(x,y)$ is non-bridge. On the other hand, if edge $(x,y)$ is a bridge, we have two sets of affected nodes $A_x$ and $A_y$, representing the nodes reachable from node $x$ and $y$, respectively. Let $G(V,E)$ be the original network, as shown in Figure \ref{fig:aff}(a), and Figure \ref{fig:aff}(b) shows the updated network $G'=G (V,E \backslash (g,h))$. When an edge $(g, h)$ is deleted from $G$, the score of nodes $\{f, g, k, l, h, i, j, m\}$ changes as shown in Figure \ref{fig:aff}(b), and these nodes are called affected nodes whereas score of the nodes $\{a, b, c, d, e\}$ does not change, and therefore, these nodes are unaffected. 
 \begin{figure}[h!]
  \centering
    \includegraphics[width=0.98\columnwidth]{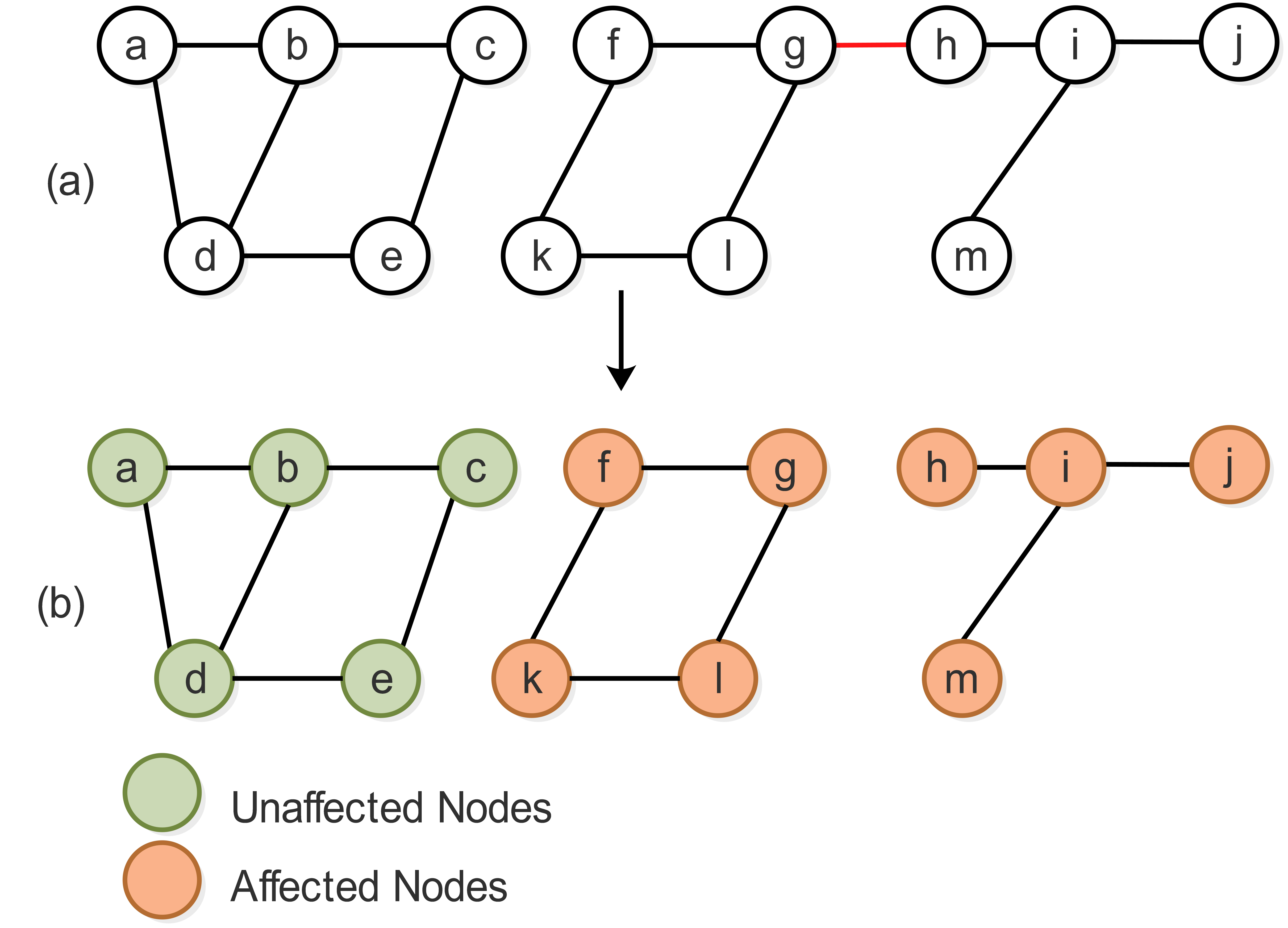}
    \caption{Illustration of affected and unaffected nodes due to updates in the network (a) Original network (b) Updated network.}
    \label{fig:aff}
  \end{figure}
\\
\textbf{Lemma 1.} Given a graph $G = (V,E)$ and an edge update $(a, b)$, any node $v \in V$ is an affected node if $u(v, a) = 1$ or $u(v, b) = 1$, resulting in $c(v)$ before deletion not equal to $c'(v)$ after deletion. Otherwise, the node is unaffected.\\Here, $c'(v)$ is the pairwise connectivity score of node $v$ in updated graph $G'$.

\subsection{Various Cases for Edge Deletion}
In this section, we discuss various cases due to the deletion of an edge from the network.

\noindent \textbf{Case 1: No change in connected component (Non-bridge edge)}

\noindent When the deleted edge $(a,b)$ is non-bridge, then there is no change in the connected components of the updated network. Node $a$ and $b$ still belong to the same connected component and the score of only few nodes in the component changes, i.e., $c(r) \neq c'(r), \, r\in C(a)$. Here, $C(a)$ represents the connected component containing node $a$.\\
\textbf{Case 2: Split connected component (Bridge edge)}

\noindent When edge $(g,h)$ is a bridge, its deletion splits the connected component into two new connected components. The score of the nodes in the component containing node $g$ and node $h$ changes, i.e., $c(r) \neq c'(r) \; \forall \,r \in C(g)$ or $C(h)$.

\subsection{Fast Computation of Pairwise Connectivity Score}

With the change in the structure of the network, it is important to update the pairwise connectivity score of the affected nodes. Let $v \in R$ be a set of nodes not reachable from node $i$. Therefore, nodes in $R$ do not contribute to the pairwise connectivity score of node $i$. Hence, it is not required to traverse the whole network to compute the score of node $i$ (as in equation 2), instead traversing the component to which node $i$ belongs is sufficient. Updated pairwise connectivity score for node $i$ can be calculated as:

\begin{equation}
c(i)=P(C(i))-P(C(i)\backslash\{i\})
\end{equation}
Here, $P(C(i))$ denotes the pairwise connectivity score of the component containing node $i$.

\subsection{Updating top-$k$ Spanners}

This section discusses the procedure for updating top-$k$ spanners. We use Algorithm \ref{alg 1} to obtain the initial spanner set and pairwise connectivity score of the nodes in the original network. In addition, we use max-heap priority queue $Q$, where the nodes are sorted by their pairwise connectivity score. Besides, we have maintained min-heap priority queue for the spanner nodes in Top-$k$.\\ 
\textbf{Lemma 2.} Given a graph $G = (V, E)$ and an edge update $(a,b)$, if the deleted edge $(a,b)$ is a non-bridge edge, then $c'(v) \geq c(v)$, $\forall \,v \in A$.\\
\textbf{Lemma 3.} Given a graph $G = (V,E)$ and an edge update $(a, b)$, if the deleted edge $(a, b)$ is a bridge edge, then $c'(v) < c(v)$, $\forall \,v \in A$.

When an edge $(a,b)$ is deleted from the network, it is first determined if it is a bridge or non-bridge edge. We then identify the set of affected nodes using the procedure discussed in Section IV(A). We compute the new pairwise connectivity score of the affected nodes using equation 4 and update the score of these nodes in the priority queue $Q$. The score of the nodes in Top-$k$ may also changes due to updates in the network, and therefore, we need to update the score of affected nodes in Top-$k$. 

Once we have updated score of all the nodes, we update the Top-$k$ spanner set. Let $w$ denotes the node with the maximum score in $Q$. Now, we compare the score of $w$ with the minimum score node in Top-$k$, and if $c(w) \leq $ Top-$k.getMin()$, we terminate the process and return Top-$k$. In contrast, if $c(w) >$ Top-$k.getMin()$, and node $w$ is already present in Top-$k$, we remove $w$ from the network and update score of the nodes in the component containing node $w$ in $Q$. On the other hand, if node $w$ is not present in Top-$k$, we remove the minimum score node from Top-$k$ and add node $w$ to Top-$k$. Finally, $w$ is removed from $G$, and the score of the nodes in the priority queue is updated. The process for updating top-$k$ spanner is repeated for a maximum of $k$ times or earlier, incase the procedure terminates.

\section{Experimental Results}
This section analyses the performance of the proposed solution. We implemented our method in Python 3.7. The experiments are performed on Window 10 PC with CPU 3.20 GHz and 16 GB RAM. We analyse the performance of the proposed solution on four real networks, Karate\footnote{\url{http://www-personal.umich.edu/~mejn/netdata/karate.zip}}, Dolphins\footnote{\url{http://www-personal.umich.edu/~mejn/netdata/dolphins.zip}}, American College Football\footnote{\url{http://www-personal.umich.edu/~mejn/netdata/football.zip}}, and HC-BIOGRID\footnote{\url{https://www.pilucrescenzi.it/wp/networks/biological/}}. The characteristics of the datasets are summarized in Table \ref{dataset1}.

\begin{table}[!h] 
\caption{Summary of real datasets.}
\label{dataset1}
\renewcommand{\arraystretch}{1.1}
\centering 
\begin{tabular}{lccc}\hlineB{2.5}
Dataset & Nodes & Edges & Avg. degree\\ \hlineB{2.5}
Karate & 34 & 78 & 4 \\ \hline 
Dolphins & 62 & 159 & 5\\ \hline 
Football & 115 & 613 & 10\\ \hline 
HC-BIOGRID & 4039 & 14342 & 7\\ \hlineB{2.5} 
\end{tabular}
\end{table}

To evaluate the  performance of the proposed solution, we compare it against static recomputation. Table \ref{result_dataset1} shows the speedup achieved by the proposed solution over recomputation. Here, speedup is the ratio of the run time of static recomputation to the dynamic solution. In order to determine how the two solutions (static and proposed dynamic solution) perform for the dynamic network, we start with a full network and randomly remove 50 edges, one at a time. We then compute the geometric mean of the speedup for the proposed solution in terms of its execution time against the static recomputation. The column Gmean, Min and Max contain geometric mean of the speedup (over 50 edge deletions), minimum speedup, and maximum speedup achieved. 

\begin{table}[!h] 
\caption{Speedup on recomputation on real datasets.}
\label{result_dataset1}
\renewcommand{\arraystretch}{1.1}
\centering
\scalebox{0.97}{%
\begin{tabular}{l|c c c|c c c} \hlineB{2.5}
Dataset & \multicolumn{3}{c}{\textbf{$k= 1$}} & \multicolumn{3}{|c}{\textbf{$k= 5$}}\\ \hline
& Gmean & Min & Max & Gmean & Min & Max \\ \hlineB{2.5}
Karate & 2.35 & 1.73 & 3.1 & 3.92 & 2.98 & 4.18 \\ \hline 
Dolphins & 3.34 &  2.11 & 4.18 & 4.16 & 3.06 & 5.33 \\ \hline 
Football & 3.72 & 3.42 & 4.21 & 10.17 & 9.6 & 11.47 \\ \hline 
HC-BIOGRID & 3.76 & 1.85 & 4.11 & 11.16 & 10.21 & 11.89 \\ \hlineB{2.5}
\textbf{Mean (Geometric)} & \textbf{3.24} & \textbf{2.19} & \textbf{3.87} & \textbf{6.56} & \textbf{5.47} & \textbf{7.42}\\ \hlineB{2.5}
\end{tabular}}
\end{table}

We run our solution for 2 different values of $k$, i.e., $k$ = 1 and 5. For real networks, the gmean speedup is always at least 2.35 for $k$ = 1 and 3.92 for $k$ = 5. The experimental results demonstrate that speedup increases with the value of $k$. The average speedup reaches 11.16 for $k$ = 5 from a speedup of 3.76 for $k$ = 1 (HC-BIOGRID dataset), which shows a significant improvement for a larger value of $k$. The average speedup over all tested datasets is 3.24 for $k$ = 1 and 6.56 for $k$ = 5. The minimum speedup achieved by the proposed solution is 1.73 for the Karate dataset, and the maximum speedup achieved is 11.89 for HC-BIOGRID dataset. In addition, it has been observed that the speedup also increases with the size of the network. For a small size network (Karate dataset), the speedup is 3.92 for $k$ = 5. In contrast, for the same value of $k$, the speedup increases significantly to 11.16 for the large size network (HC-BIOGRID dataset).

\section{Conclusion}
This paper studied the SH spanner identification problem for the dynamic networks, namely SST problem. We proposed an efficient solution for SST problem that maintains top-$k$ spanners dynamically by identifying the set of affected nodes whose pairwise connectivity score changes due to updates in the network. We also designed a fast mechanism for computing the pairwise connectivity score of the nodes. We analysed the performance of our solution experimentally and showed that the proposed single edge update solution speedup the process by a minimum factor of 3.24 over recomputation.

\section{ACKNOWLEDGMENTS}
This work was completed under Australian Government Research Training Program Scholarship at the University of Adelaide. It is supported by Key-Area Research and Development Plan of Guangdong Province \#2020B010164003. The corresponding author is Hong Shen.

\bibliographystyle{IEEEtran}
\bibliography{newref}
\end{document}